\documentclass[osajnl,a4paper,twocolumn]{revtex4}  

\usepackage[latin1]{inputenc}
\usepackage{amstext}

\begin{document}

\title{Quasi-Optical Characterization of Dielectric and Ferrite Materials}

\author{M. Goy, F. Caroopen,  M. Gross}

\affiliation{AB Millimetre, 52 rue Lhomond 75005 Paris: France\\
tel: +33 1 47 07 71 00, fax: +33 1 47 07 70 71,
abmillimetre@wanadoo.fr }

\author{R.I. Hunter and  G.M. Smith}

\affiliation{Millimetre Wave and High Field ESR Group, University of
St Andrews\\
 North Haugh, St Andrews, Fife KY 16 9SS, Scotland,
United Kingdom\\
       tel: +44 1334 463156, 2669, fax: 463104, rihl, gms@st-and.ac.uk }



%

\maketitle

17th International Symposium on Space THz Technology, Paris 2006 May
10-12

\section{Introduction}

\begin{figure}[]
\centering
\includegraphics[width=8.5cm]{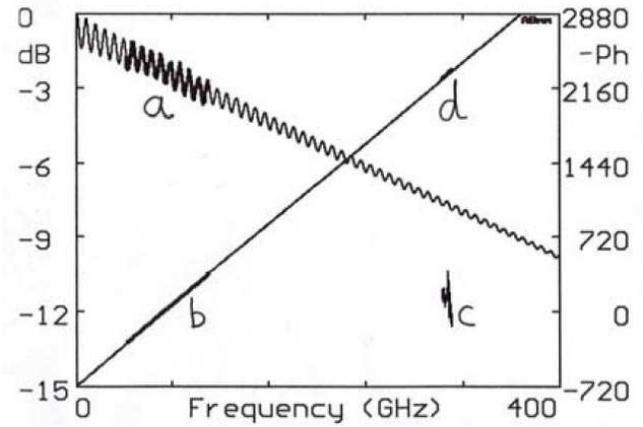}
\caption{  Transmission through 10.95 mm nylon. (a) thick line is
the observed amplitude in V-W bands. (b) is the corresponding
phase (represented in opposite sense for clarity). Around 305 GHz,
the measured (d) phase value, in good alignment with extrapolated
(b), shows that the permittivity $\varepsilon'=3.037 $ is constant
with frequency. On the contrary, the position of the amplitude (c)
shows that the loss, which was $\tan \delta = 0.013$ at low
frequencies, has increased to $\tan \delta = 0.019$, since (c) is
far from the extrapolated (a). } \label{FIG1}
\end{figure}

In the millimeter-submillimeter range, Quasi-Optical (QO) benches
can be relatively compact, typically of order 10cm wide and 1m
long. The focussing elements used in these benches are dielectric
lenses, or off-axis elliptical mirrors. Simultaneous Transmission
$T$ (corresponding to the complex $S21$ parameter), and Reflection
$R$ (corresponding to the complex S ii parameter) are vectorially
detected versus frequency in the frequency range 40-700 GHz. A
parallel-faced slab, thickness $e$, of dielectric material is
placed at a Gaussian beam waist within the system. It is
straightforward to determine the refractive index $n$ (with
$\varepsilon '= n^2$) of this sample from the phase rotation
$\Delta \Phi$:
\begin{equation}\label{equ1}
    (n-1) e /\lambda=\Delta
\Phi/360
\end{equation}
The loss factor $\tan \delta  $ is known from the damping of the
transmitted signal, Fig. \ref{FIG1}:
\begin{equation}\label{Eq2}
   \tan \delta  = 1. 1 ~\alpha \textrm{(dB/cm)}/n F \textrm{(GHz)}
\end{equation}

\begin{figure}[]
\centering
\includegraphics[width=8.5cm]{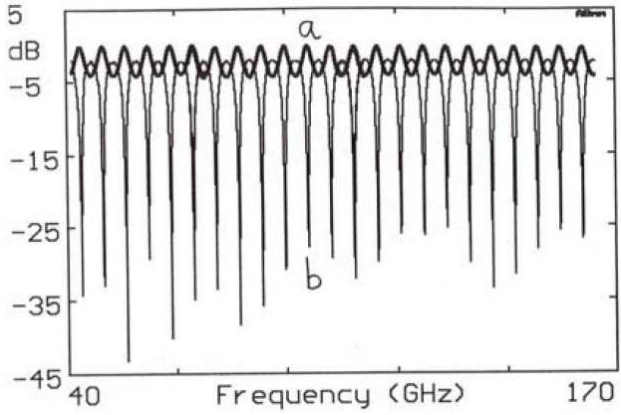}
\caption{Transmission (a) and reflection (b) through 9.53 mm AIN.
The dielectric parameters, observed in V-W-D bands, are constant
with $ \varepsilon'=8.47S$, $\tan \delta = 0.0007$, also measured
the same in cavity at 140.4 GHz.} \label{FIG2}
\end{figure}

\begin{figure}[]
\centering
\includegraphics[width=8.5cm]{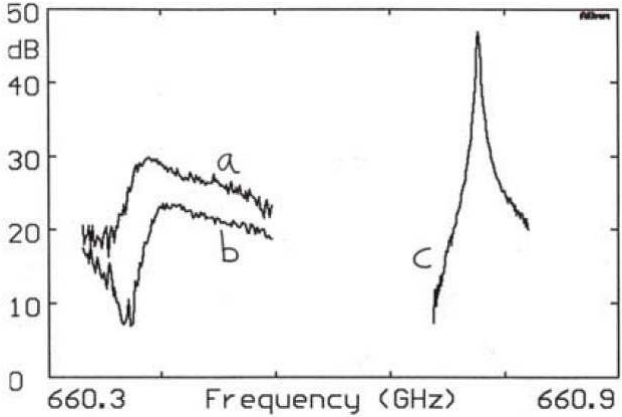}
\caption{ Resonances observed in an open Fabry-Perot cavity loaded
by a 10.03 mm thick slab of slightly birefringent Teflon. In (a)
the RF field is along the large index axis  $ \varepsilon ' =
2.0664$, in (b) along the small index axis $ \varepsilon
'=2.0636$. In (c), is the empty cavity resonance. The $\varepsilon
'$ anisotropy is exactly the same as observed at 135 GHz. The loss
has increased from $\tan \delta =0.0003$ at  135 GHz to $0.0008$
at 660 GHz. } \label{FIG3}
\end{figure}

The samples in this measurement system act as Fabry-Perot
resonators with maximum transmission corresponding to minimum
reflection, and vice-versa (see Fig.\ref{FIG2}), with a period
$\Delta F=c/(2 n e)$. For very low loss materials, there is
however some difficulty in measuring the loss term by a single
crossing, since the maximum transmission is very close to 0 dB.
One uses the cavity perturbation technique, which makes visible
the low losses after many crossings through the dielectric slab
(see Fig.3).

\section{Experimental Setup for free-space
 propagation}

In V-W-D bands (extended down to ca 41 GHz, close to the V-band
cutoff), we use the following waveguide components. On the source
side, the harmonic Generator HG sends its millimeter power through
a full-band Faraday isolator FI1, cascaded with a fixed attenuator
AT1, a directional coupler DC (from port 2 to port 1) and a Scalar
Horn SH1. The reflection (Channel 1) is detected by a Harmonie
Mixer HM1 attached to output 3 of the DC through the isolator F12.
On the transmission detection side, the Scalar Horn SH2 sends the
collected wave to the Harmonic Mixer HM2 (Channel 2) through
cascaded AT2 and FI3.

\section{Isolators $\textrm{FIs}$ and Attenuators $\textrm{ATs}$, what
for?}\label{section3}

\begin{figure}[]
\centering
\includegraphics[width=8.5cm]{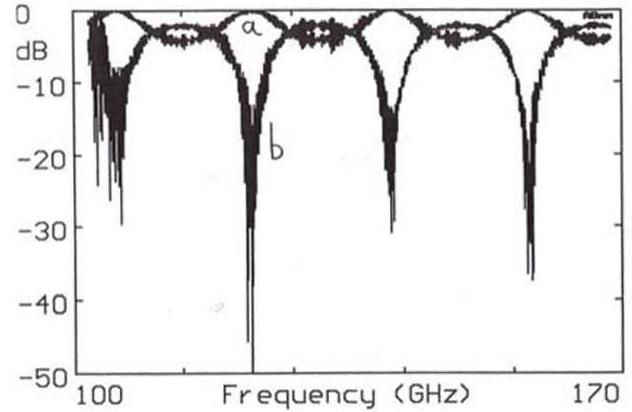}
\caption{Transmission (a) and reflection (b) through a 3~mm thick
$MgAl_2O_3$ slab. These raw data show parasitic standing waves
appearing as noise } \label{FIG4}
\end{figure}

The first use of isolators is to assume a one-way propagation. The
non-linear devices HG and HMs contain Schottly diodes. In case the
wave can travel go-and-back from one device to the other, the
combination of non-linear and standing waves effect can send
microwave power from a given harmonic to another harmonic
\cite{goy_1998_sand} . This is why multipliers cascaded without
isolation (like $\times 2 ~\times 3$ ) can create unexpected
harmonics (like $\times 5 ~\times 7$). We have also observed, for
instance with cascaded tripiers ($ \times 3 ~\times 3 $ ),
measurable amounts of unexpected $\times 10,~ \times 11$, or $
\times 12 $ \cite{goy_2005_priv}. The devices HG and HMs can be
viewed as Schottky diodes across waveguides, meaning unmatched
structures. The second use of the FIs is to reduced the VSWR.
Their typical return is -20~dB (VSWR \emph{ca} 1.22). We improve
this value down to - 30~dB (VSWR \emph{ca} 1.07) when introducing
the fixed attenuators ATs.

\section{Experimental difficulties}

\begin{figure}[]
\centering
\includegraphics[width=6cm]{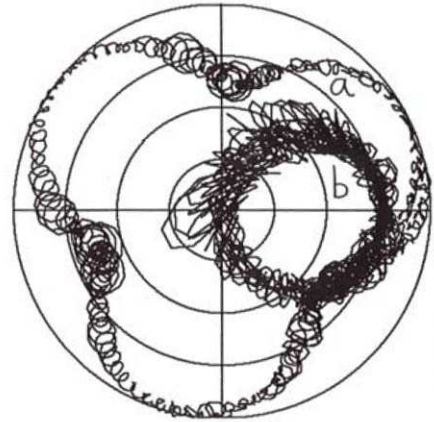}
\caption{Polar plot of Fig.\ref{FIG5} } \label{FIG5}
\end{figure}

\begin{figure}[]
\centering
\includegraphics[width=8.5cm]{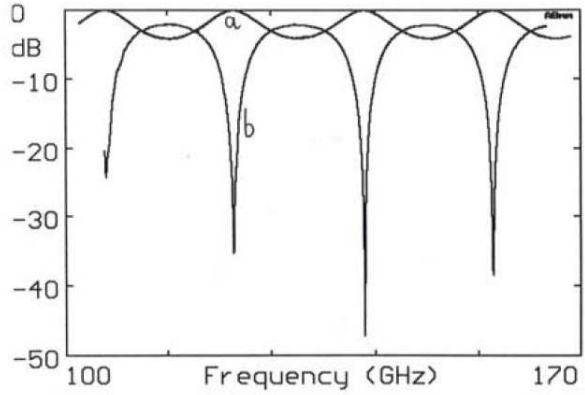}
\caption{Same as Fig.\ref{FIG4} after FT filtering.. The measured
dielectric parameters are $\varepsilon '= 8.080$, and $\tan \delta
= 0.0005$. } \label{FIG6}
\end{figure}

\begin{figure}[]
\centering
\includegraphics[width=6cm]{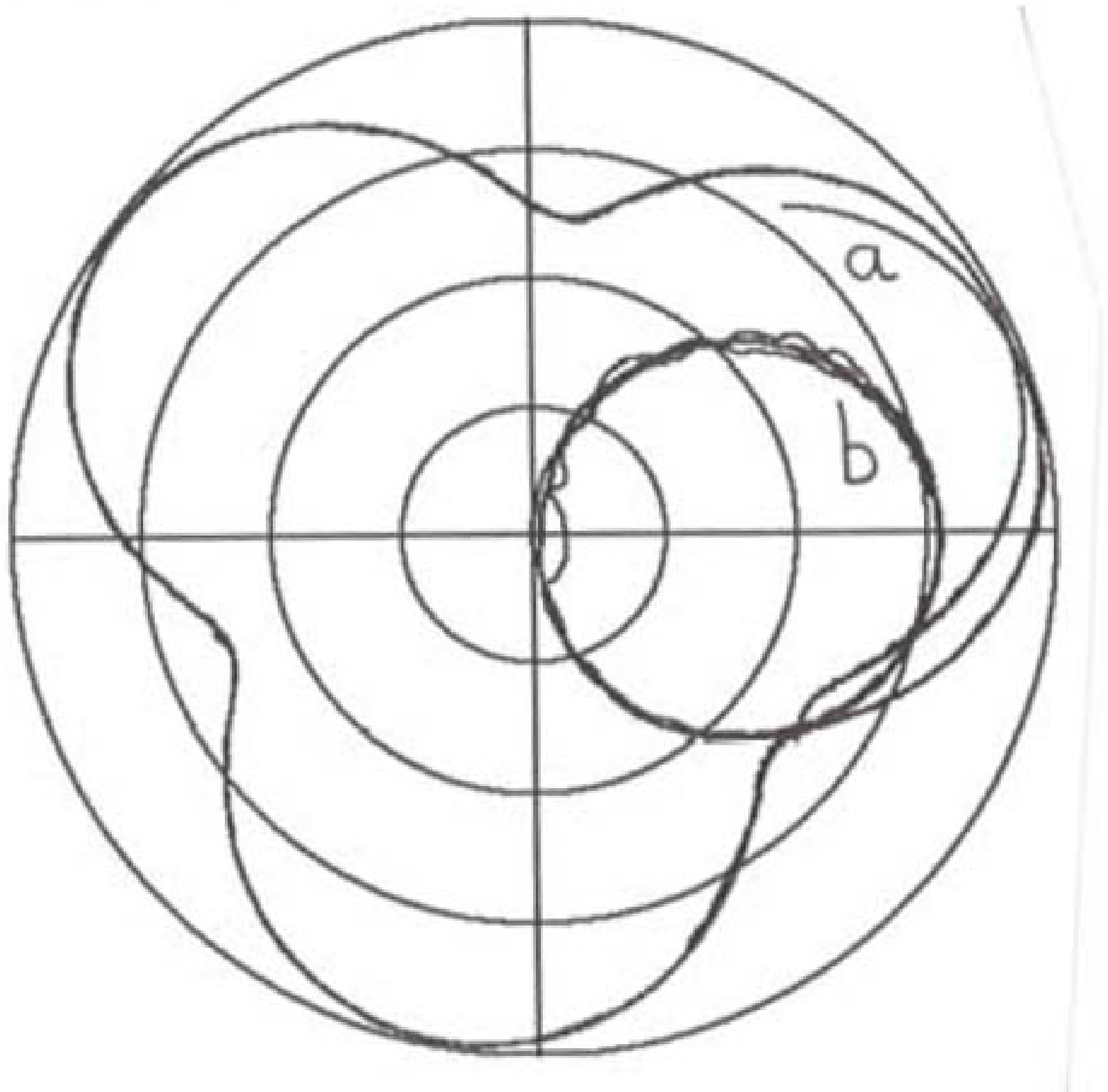}
\caption{Polar plot of Fig.\ref{FIG6} } \label{FIG7}
\end{figure}

\begin{figure}[]
\centering
\includegraphics[width=6cm]{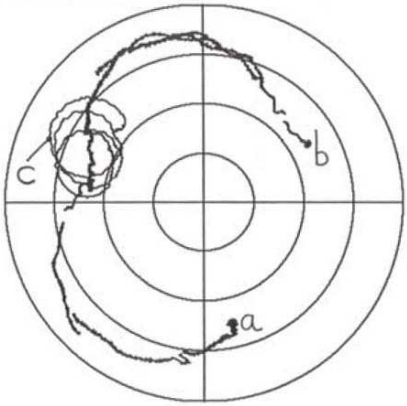}
\caption{Transmission across 9.97~mm sapphire, from 469~GHz, point
(a), to 479~GHz, point (b), with absorbers, total 30~dB, between
source and detection, in order to reduce the parasitic standing
waves. Despite this strong attenuation, the measurement quality is
far from being as good as at lower frequencies, like in
Fig.\ref{FIG7}. In (c) is the 473.3 to 474.5~GHz sweep without
absorbers, showing big standing waves effects.} \label{FIG8}
\end{figure}

Even with our best benches using the complete chains assuming a
low VSWR ( $<1.1$ see sec \ref{section3}), the parasitic standing
waves effects are clearly visible on raw data,
Fig.\ref{FIG4}-\ref{FIG5}. They are due to multiple reflections
between the sample, placed perpendicular to the beam, and the
components of the bench. However, they can be completely filtered
by FT calculations (see Figs.\ref{FIG6}-\ref{FIG7}). There is a
lack of FIs waveguide isolators above 220~GHz and, as far as we
know, of DCs above 400~GHz. As a consequence, characterization at
submillimeter wavelengths is operated by transmission only, and is
much more difficult, Fig.\ref{FIG8}, than in V-W-D-bands, due to
large parasitic standing waves.

\section{Non-magnetized ferrites characterization}

\begin{figure}[]
\centering
\includegraphics[width=8.5cm]{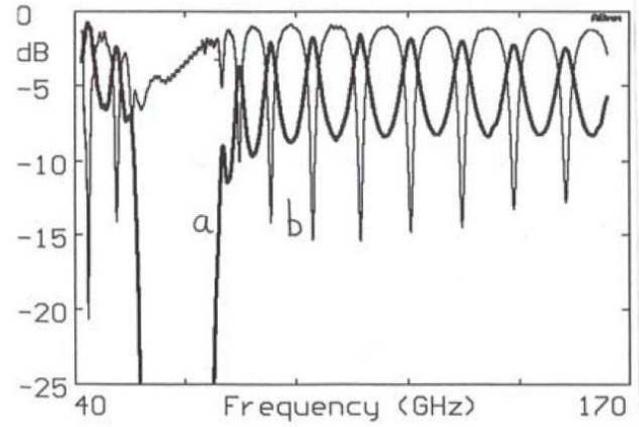}
\caption{Transmission (a), and reflection (b)  through a  2.55 mm
thick non-magnetized TDK ferrite sample. } \label{FIG9}
\end{figure}

In the case of ferrite materials, the properties are very strongly
frequency dependent. Non-magnetized ferrites show a strong
resonance in the range 50-60~GHz (see Fig.\ref{FIG9}), and the
asymptotic behavior, far from resonance, starts to be visible
beyond 200~GHz. Measurements performed at 475~GHz on six samples
give $\varepsilon'$ in the range 18.8 to 21.4, and $\tan \delta$
in the range 0.012 to 0.018.

\section{Magnetized ferrites}

\begin{figure}[]
\centering
\includegraphics[width=8.5cm]{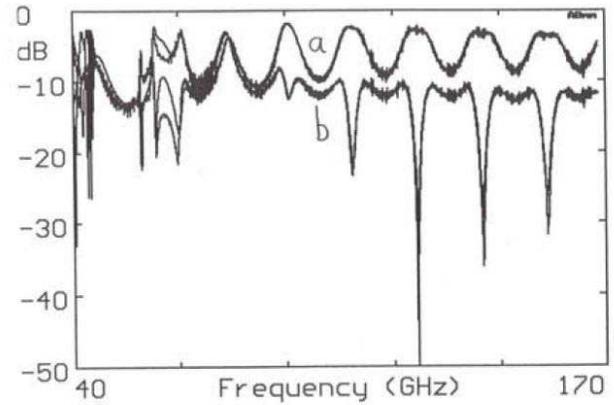}
\caption{Transmission  through  2 mm  magnetized sample FB6H1, (a)
is $-45^\circ$, (b) is $+45^\circ$. Experimental traces and
superimposed fittings. } \label{FIG10}
\end{figure}

\begin{figure}[]
\centering
\includegraphics[width=8.5cm]{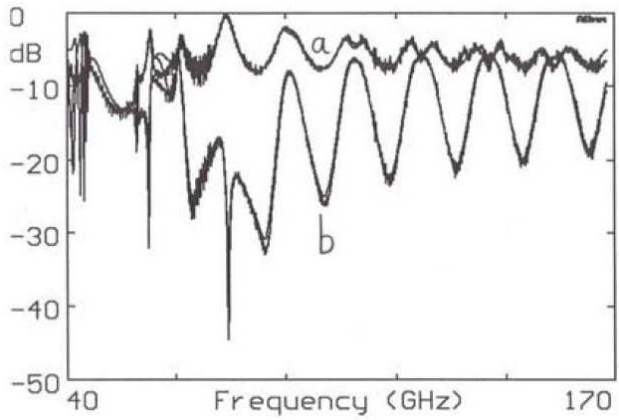}
\caption{ Same as Fig.\ref{FIG10}, where (a) is $90^\circ$ and (b)
is $0^\circ$.} \label{FIG11}
\end{figure}

\begin{figure}[]
\centering
\includegraphics[width=8.5cm]{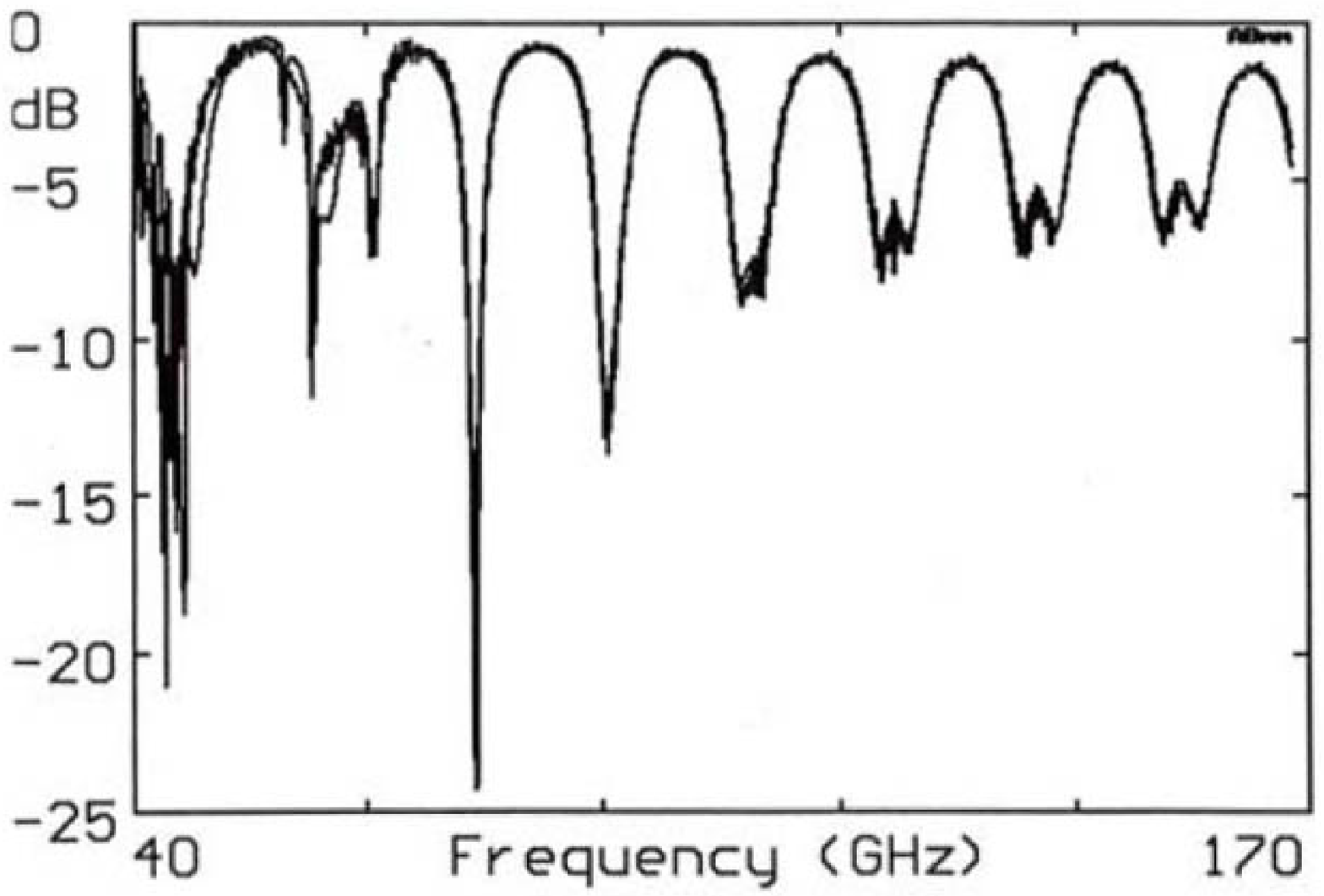}
\caption{Reflection at $0^\circ$ from the magnetized sample FB6H1,
experiment and fit. } \label{FIG12}
\end{figure}

When a ferrite is submitted to an external, or internal, magnetic
field, there is a strong anisotropy of propagation according to the
circular polarization of the  crossing  electromagnetic  wave
\cite{cite_3}. The  two refractive indices $n^\pm$ are given by:
$$
(n^\pm)^2 = \varepsilon ' [1 + F_m / (F_0 \pm F) ]
$$
, where $F$ is the frequency, $F_0$ the Larmor frequency, $F_m$ is
proportional to the remanent magnetization of the ferrite, and
$\varepsilon '$ is the dielectric constant. Any  linearly polarized
wave, like ours at the SH outputs, can be viewed as the
superposition of two opposite senses   circularly  polarized
components.   After crossing the ferrite,  one of the components has
experienced a larger retardation than the other, so that, when
recombining the two, the plane of linear polarization has been
rotated. In order to characterize magnetized ferrite samples, it is
necessary to measure not only the transmitted signals with a
polarization parallel to the source, but also those with
polarization at $ \pm 45$ degrees, and 90 degrees, see
Figs.\ref{FIG10}-\ref{FIG11}-\ref{FIG12}.

\begin{figure}[]
\centering
\includegraphics[width=8.5cm]{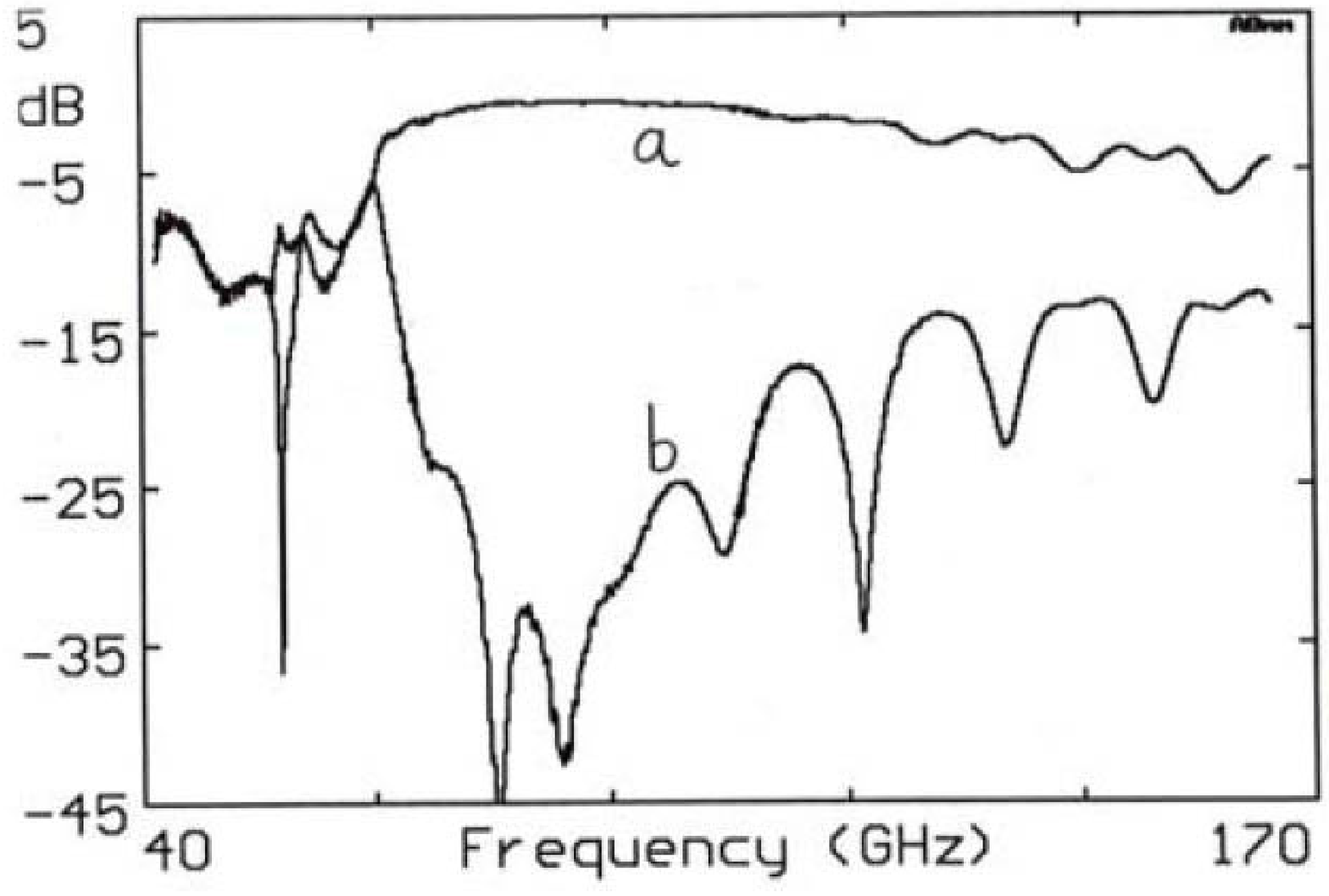}
\caption{Reflection at $0^\circ$ from the magnetized sample FB6H1,
experiment and fit. } \label{FIG13}
\end{figure}

\begin{figure}[]
\centering
\includegraphics[width=8.5cm]{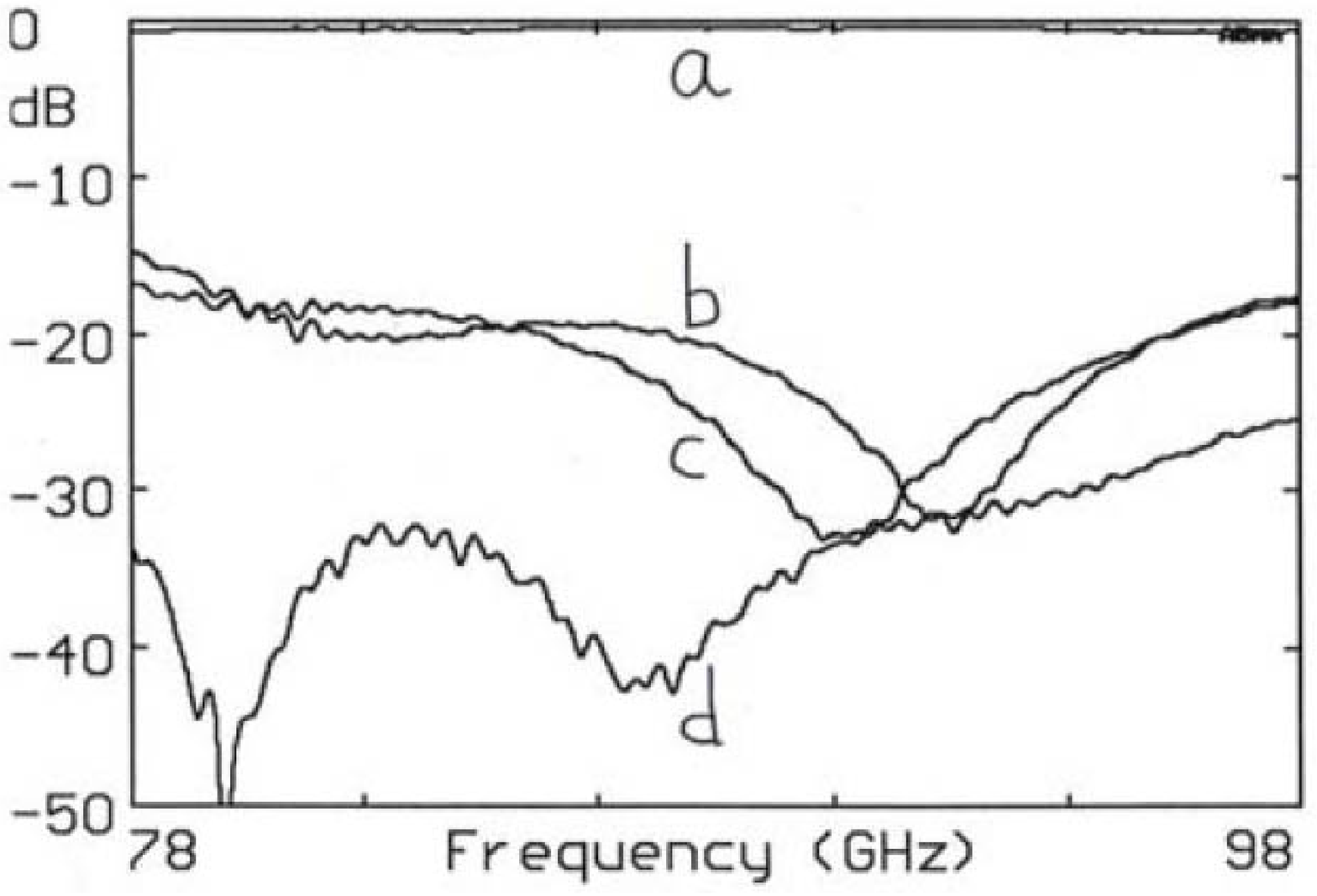}
\caption{Reflection at $0^\circ$ from the magnetized sample FB6H1,
experiment and fit. } \label{FIG14}
\end{figure}

\begin{figure}[]
\centering
\includegraphics[width=8.5cm]{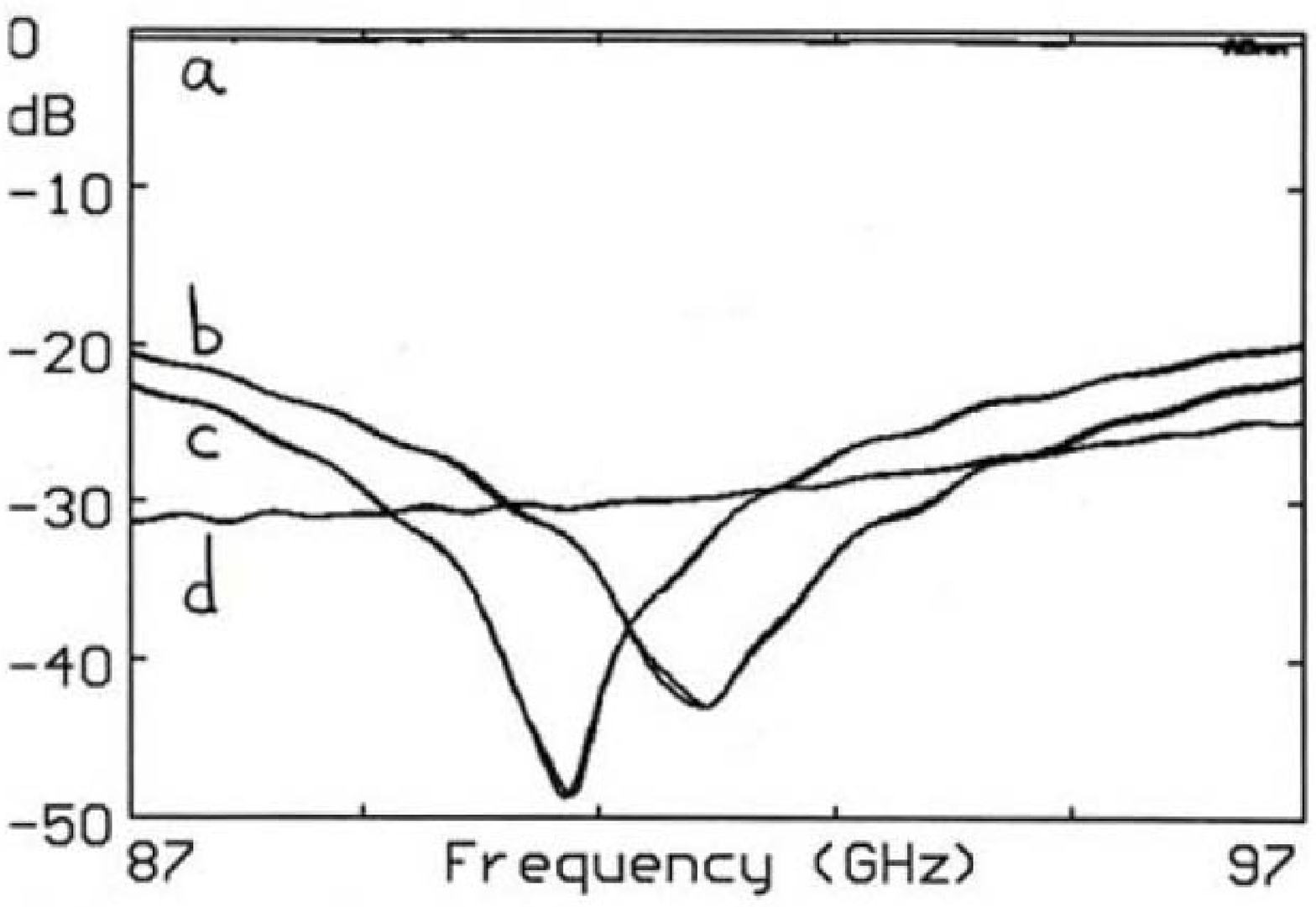}
\caption{Reflection at $0^\circ$ from the magnetized sample FB6H1,
experiment and fit. } \label{FIG15}
\end{figure}

When adding an anti-reflection coating on each side of the
magnetized ferrite, the thickness of the ferrite being chosen so
that the rotation through it is $45^\circ$ at the required
frequency, one can obtain a good QO Faraday Rotator,
Fig.\ref{FIG13}. The performances observed around the central
frequency, Figs.\ref{FIG14}-\ref{FIG15}, are at least similar (for
isolation or matching) or better (for insertion loss) than the
equivalent waveguide isolators.

\section{$\textrm{QOFRs}$ expected to become submillimeter isolators and directional Couplers}

\begin{figure}[]
\centering
\includegraphics[width=8.5cm]{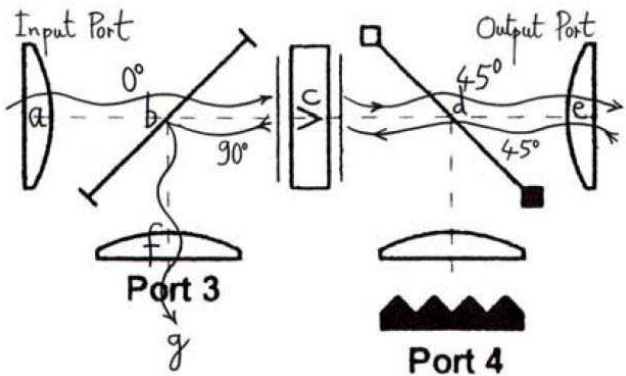}
\caption{ Schematic diagram of a QOFR used as an isolator (there
is a inatched load at (g) ) Port 3, or as a Directional Coupler DC
for detecting reflected waves at (g). The vertical polarisation at
(a), fully transmitted through the horizontal grid (b), rotates by
$+45^\circ$ through the magnetized ferrite (c), then is fully
transmitted through the $-45^\circ$ grid (d). Any reflected signal
without polarisation change will cross back (d) without loss, then
will rotate by $+ 45^\circ$ again across (c), becoming horizontal,
then will be totally reflected by the horizontal grid (b), towards
Port 3. } \label{FIG16}
\end{figure}

Our QO benches studying samples perpendicular to the wave beam,
are, up to now, less performing  in submillimeter (Fig.\ref{FIG8})
than in the millimeter domain (Fig.\ref{FIG7}), due to parasitic
standing waves. Introducing the appropriate QO Faraday Rotators
will reduce this effect. On figure \ref{FIG16} one can see how a
QOFR can be simply configurated for that purpose.

\section{Conclusion}

Precise and quick QO measurements in the 40-170~GHz interval, in
particular for ferrites characterization, opens the possibility of
similar precise and easy measurements at high frequencies, including
the submillimeter domain, by using these ferrites in QOFRs in
progress \cite{cite_4}. At the same time, widely sweepable
solid-state submillimeter sources must be developed.
%
%
%
%
%


\end{document}